# Effect of Mg-Al insertion on magnetotransport properties in epitaxial Fe/sputter-deposited MgAl$_2$O$_4$/Fe(001) magnetic tunnel junctions

Mohamed Belmoubarik, Hiroaki Sukegawa, Tadakatsu Ohkubo, Seiji Mitani,

and Kazuhiro Hono

National Institute for Materials Science, 1-2-1 Sengen, Tsukuba 305-0047, Japan

E-mail: Belmoubarik.Mohamed@nims.go.jp

(08, Nov. 2016)

## Abstract

We investigated the effect of an Mg-Al layer insertion at the bottom interface of epitaxial Fe/MgAl$_2$O$_4$/Fe(001) magnetic tunnel junctions (MTJs) on their spin-dependent transport properties. The tunnel magnetoresistance (TMR) ratio and differential conductance spectra for the parallel magnetic configuration exhibited clear dependence on the inserted Mg-Al thickness. A slight Mg-Al insertion (thickness < 0.1 nm) was effective for obtaining a large TMR ratio above 200% at room temperature and observing a distinct local minimum structure in conductance spectra. In contrast, thicker Mg-Al (> 0.2 nm) induced a reduction of TMR ratios and featureless conductance spectra, indicating a degradation of the bottom-Fe/MgAl$_2$O$_4$ interface. Therefore, a minimal Mg-Al insertion was found to be effective to maximize the TMR ratio for a sputtered MgAl$_2$O$_4$-based MTJ.

1. Introduction

Obtaining a large tunnel magnetoresistance (TMR) ratio exceeding 100% at room temperature (RT) in a magnetic tunnel junction (MTJ) is very essential to the development of various spintronic devices such as magnetoresistive random access memories, magnetic logic circuits and high sensitive magnetic sensors.[1] A crystalline MgO with (001) orientation is commonly used as a tunnel barrier of MTJs to attain large TMR ratios.[2–4] However, there could be a limitation of MgO material as a barrier in MTJs when ferromagnetic layers other than CoFe (or CoFeB) [such as Co-based Heusler alloys[5,6], tetragonal Mn-based (MnGe and MnGa) alloys [7,8] and FePt alloys[9,10]] are used for MTJs. The large lattice mismatches between these electrode materials and MgO (5–8%) accompanied with misfit dislocations make it difficult to fabricate MTJ structures having crystallographically coherent interfaces and exhibiting large TMR ratios at RT.[11,12]

Spinel $MgAl_2O_4$(001) barrier-based MTJs have shown large TMR ratios over 100% at RT due to the occurrence of spin-dependent coherent tunneling similar to MgO(001) barriers.[13–15] Importantly, the lattice constant of an $MgAl_2O_4$ barrier prepared by a post-oxidation method can be tuned by the Mg-Al composition in $MgAl_2O_4$, leading to nearly a perfect lattice-matching with bcc Co-Fe alloys[13,14] and $Co_2FeAl$ Heusler alloy.[15] Recently, using Fe as electrodes we developed a lattice-matched and very flat $MgAl_2O_4$(001) barrier with a cation-disordered spinel structure by a direct rf sputtering of a $MgAl_2O_4$ sintered target, instead of the conventional post-oxidation of Mg-Al alloys.[16] The TMR ratio of a direct-sputtered epitaxial

Fe/MgAl$_2$O$_4$/Fe(001) MTJ reached 245% at RT, which is larger than typical TMR ratios in epitaxial Fe/MgO/Fe MTJs.[3,17] In general, spin-dependent transport properties of an MTJ is significantly affected by the barrier/ferromagnetic interface conditions. Therefore, TMR ratio can be drastically enhanced or reduced only by a minor modification of these interfaces states; such as the insertion of an ultrathin metallic layer.[18–20] In our previous report, ultrathin Mg-Al (Mg$_{10}$Al$_{90}$ atomic%) was inserted between the bottom-Fe electrode and direct-sputtered MgAl$_2$O$_4$ barrier of Fe/MgAl$_2$O$_4$/Fe MTJs in order to tune the interface condition.[16] However, a systematic study on the Mg-Al insertion effect has yet to be investigated. Therefore, in this study, we focused on the Mg-Al insertion thickness dependence of the spin-dependent transport. We found that the Mg-Al insertion modified the MTJs properties such as, the zero bias TMR ratio, resistance-area ($RA$) product and its bias voltage dependence. Especially, an Mg-Al thickness more than 0.4 nm inserted under a 1.9-nm-thick MgAl$_2$O$_4$ barrier significantly reduced the TMR ratio. The differential conductance spectra also suggested the reduction of the degree of the coherent tunneling transport through the barrier when the Mg-Al insertion thickness increased. Thus, the interface modification by Mg-Al insertion under the sputtered MgAl$_2$O$_4$ barrier formed an insufficient oxidation interface, which weakened the spin-dependent coherent tunneling effect. On the other hand, a minimal insertion (less than 0.1 nm) was effective to prevent the interface oxidation at the bottom-Fe and enhance the TMR ratio to maximum values (240–245%).

## 2. Experimental details

The MTJ multilayers were fabricated using a dc and rf magnetron sputtering system with a base pressure of $6 \times 10^{-7}$ Pa. The MTJ stack consisted of an MgO(001) substrate/Cr (40)/Fe (100)/Mg$_{10}$Al$_{90}$ (Mg-Al, $t_{MgAl}$ = 0–0.6)/MgAl$_2$O$_4$ (1.9)/Fe (7)/Ir$_{20}$Mn$_{80}$ (12)/Ru (10) (unit in nm). Here, $t_{MgAl}$ is the nominal thickness of a wedge-shaped Mg-Al insertion formed by a linear motion shutter. The MgAl$_2$O$_4$ barrier was directly deposited from a stoichiometric MgAl$_2$O$_4$ sintered target. Each layer of the MTJ stack was deposited at RT, then was succeeded by an *in-situ* post-annealing at temperatures mentioned in Fig. 1. Here, the deposition and post-annealing conditions were optimized using X-ray and reflection high-energy electron diffractions (see Ref 16). Meanwhile, we confirmed that on an Mg-Al insertion layer ($t_{MgAl}$ up to 0.6 nm) MgAl$_2$O$_4$ layers grew epitaxially and had the cation-disordered structure. This is necessary for obtaining large TMR ratios in spinel MgAl$_2$O$_4$-based MTJs with Co-Fe based electrodes.[13,21] Then, the MTJ stack was patterned into elliptical pillars with a dimension of 5×10 µm$^2$ using photolithography and Ar ion-beam etching. The MTJ devices were characterized using a conventional dc 4-probe method with an external magnetic field directed along Fe[100]. Here, a positive bias voltage indicated electrons tunneling from the top Fe electrode to the bottom one. All presented data in this report were measured at RT.

## 3. Results and discussion

The $t_{MgAl}$ dependence of the TMR ratio and $RA$ in the parallel (P) state ($RA_P$) of epitaxial Fe/Mg-Al($t_{MgAl}$)/MgAl$_2$O$_4$ (1.9 nm)/Fe(001) MTJs are illustrated in Figure 2 (a). A bias voltage less than 10 mV was used for these measurements. Near the wafer edge, i.e. $t_{MgAl}$ < 0.05 nm, the largest TMR ratio of 245% was observed closely to that reported in our previous report ($t_{MgAl} \approx 0.02$ nm).[16] Therefore, the bottom-Fe/MgAl$_2$O$_4$ interface smoothness, oxidation state, and high crystallinity were conserved for this thickness region. As increasing $t_{MgAl}$, the TMR ratio basically decreased and the $RA_P$ increased. We can distinguish three main regions for the TMR ratio and $RA_P$ changes. When $t_{MgAl}$ is less than 0.1 nm (region-I), the TMR ratio is greater than 200% ($RA_P$: 6–8 k$\Omega\cdot\mu$m$^2$). The second region (region-II); for which $t_{MgAl}$ is in the range of 0.13–0.36 nm; exhibited a TMR ratio and $RA_P$ in the ranges of 170–190% and 14–17 k$\Omega\cdot\mu$m$^2$, respectively. For the third region (region-III); where $t_{MgAl}$ is above 0.44 nm; the TMR ratio rapidly dropped to less than 100% and the $RA_P$ increased to be within 38–43 k$\Omega\cdot\mu$m$^2$. Three representative TMR curves of these regions are illustrated in Figure 2 (b). Obviously, the large TMR ratios and low $RA_P$ of region-I indicated the dominance of coherent spin-dependent tunneling across MgAl$_2$O$_4$ barriers, as discussed later. The $t_{MgAl}$ increase accompanied by an $RA_P$ increase is attributed to an increase of the effective barrier thickness by an oxidation of the Mg-Al insertion layer. This oxidation can occur during the MgAl$_2$O$_4$ sputter-deposition and/or the subsequent post-annealing process (at 500°C for 15 min). This means that oxygen deficiencies are induced near the bottom-Fe interface for a thick Mg-Al insertion layer. Consequently, this led to a lowering of the barrier crystallinity and a formation of oxygen vacancies similarly to epitaxial Fe/MgO/Fe MTJs.[22] Interestingly, in the region-II, the TMR ratio and $RA_P$ did not show strong dependence on $t_{MgAl}$, which suggested that the coherent transport can be maintained even in the presence of a small amount of oxygen deficiencies.

For more insight into the effect of the Mg-Al insertion on spin-dependent tunneling properties, we evaluated the bias voltage ($V$) dependence of the TMR ratio for various $t_{MgAl}$ as shown in Fig. 3 (a). Here, the TMR ratio was normalized to the TMR ratio peak. Consistently with the tendency of the TMR ratio under a nearly zero bias (Fig. 2 (a)), as $t_{MgAl}$ increased the bias voltage dependence became stronger. For a thick $t_{MgAl}$, the peak position of the TMR ratio shifted to the positive bias direction (see the curve of $t_{MgAl} = 0.55$ nm). These behaviors indicate the degradation of the interface state of $MgAl_2O_4$ barrier by a thick Mg-Al insertion. To quantify the bias voltage dependence, we introduced the parameter $V_{half (+)}$ ($V_{half (-)}$) defined as the positive (negative) $V$ value where the TMR ratio is reduced to the half of its zero bias value. The extracted $V_{half (+)}$ and $V_{half (-)}$ as a function of $t_{MgAl}$ are shown in Fig. 3 (b). For the region-I ($t_{MgAl} < 0.1$ nm), the high values of $V_{half (+)} = 1.2$ V and $|V_{half (-)}| = 1.0$ V were obtained. The $V_{half}$ was reported to be directly correlated with the barrier oxidation degree for Fe/natural or plasma post-oxidized Mg-Al-O/Fe MTJs, which exhibited high $V_{half}$ values (~1.2 V) as well as high zero bias TMR ratios (~212%) for the optimized oxidation conditions.[14] Therefore, the sputtered $MgAl_2O_4$ barrier for the region-I also had an adequate interface oxidation and chemically sharp bottom- and top-barrier interfaces even though the Mg-Al insertion thickness is less than one atomic monolayer ($t_{MgAl} < 0.1$ nm). Interestingly, $t_{MgAl}$ dependence of $V_{half\,(+)}$ (Fig. 3 (b)) and the TMR ratio (Fig. 2 (a)) exhibited the same decreasing tendency by three steps corresponding to regions I, II and III. This change reflected a gradual degradation of the barrier oxidation (interface) state by the simultaneous formation of oxygen vacancies in the barrier and the oxidation of Mg-Al insertion layers for regions II and III. Note that the small asymmetry with respect to the bias direction is attributed to the slight structural difference between the bottom- and top-Fe/MgAl$_2$O$_4$ interfaces induced by our multilayer deposition conditions.

For further investigations on the TMR degradation mechanism with the $t_{MgAl}$ increase, we plotted the normalized differential conductance ($G = dI/dV$) in the parallel and antiparallel (AP) states (denoted as $G_P$ and $G_{AP}$, respectively) for various $t_{MgAl}$ as shown in Fig. 4 (a) and (b). The spectra were calculated from measured current ($I$)-voltage ($V$) characteristics. In the P state, local minima around ±230 mV were clearly observed for smaller $t_{MgAl} < 0.33$ nm (region-I~II), reflecting the strong contribution of the coherent tunneling between the two Fe electrodes, and consistently with the obtained large TMR ratios.[14,16] This may be attributed to the majority $\Delta_5$-band edge of Fe.[14,16,23] While increasing $t_{MgAl}$, the local minima disappeared and the spectra became featureless. Also, the degree of asymmetry increased and finally almost parabolic shapes were obtained for $t_{MgAl} \geq 0.38$ nm (region-II~III). These features can be correlated to the increase of the incoherent tunneling contribution, which is consistent with the reduction of the TMR ratio and $V_{half}$ as observed in the post-oxidized Mg-Al-O-based MTJs.[14,23] In the AP state (Fig. 4 (b)), the spectra were almost symmetric without any local minimum structure and exhibited stronger bias voltage dependence for larger $t_{MgAl}$. This suggests that the inelastic scattering at the barrier interface and the contribution of the incoherent tunneling became significant. In the inset of Fig. 4 (b), we showed the $dG_{AP}/dV$ spectra for selected three $t_{MgAl}$. The peaks around ±60 mV reflected the interfacial magnon excitations.[14,23] The observed peak shift to a higher position for a higher $t_{MgAl}$ may be related to the degradation of the bottom-Fe/MgAl$_2$O$_4$ interface. At the end, this study showed that the adequate control of the barrier interface state is crucial for enhancing the contribution of the coherent tunneling, and thus achieving large TMR ratios in MgAl$_2$O$_4$-based MTJs.

## 4. Conclusions

The effects of Mg-Al thickness ($t_{MgAl}$) on the TMR ratio and differential conductance of (001)-oriented Fe/Mg-Al insertion/MgAl$_2$O$_4$/Fe MTJs were investigated. The dominance of the coherent tunneling through MgAl$_2$O$_4$ barriers was found to be feasible for $t_{MgAl}$ < 0.1 nm. Increasing $t_{MgAl}$ induced a degradation of the barrier quality, and hence increased the contribution of incoherent tunneling through MgAl$_2$O$_4$ tunnel barriers. Therefore, the insertion of a very thin Mg-Al layer (< 0.1 nm) was found to be quite effective for the enhancement of spin-dependent properties through sputter-formed MgAl$_2$O$_4$ barriers. The present result shows the sensitivity of the interface states of MgAl$_2$O$_4$-based MTJs to the metallic layer insertion, and the importance of the barrier-interface engineering toward practical applications of MgAl$_2$O$_4$- based MTJs.

**Acknowledgement**

This work was supported by ImPACT Program of Council for Science, Technology and Innovation.

# References

[1] S. Yuasa and D.D. Djayaprawira, J. Phys. Appl. Phys. **40**, R337 (2007).

[2] S.S.P. Parkin, C. Kaiser, A. Panchula, P.M. Rice, B. Hughes, M. Samant, and S.-H. Yang, Nat. Mater. **3**, 862 (2004).

[3] S. Yuasa, T. Nagahama, A. Fukushima, Y. Suzuki, and K. Ando, Nat. Mater. **3**, 868 (2004).

[4] S. Ikeda, J. Hayakawa, Y. Ashizawa, Y.M. Lee, K. Miura, H. Hasegawa, M. Tsunoda, F. Matsukura, and H. Ohno, Appl. Phys. Lett. **93**, 082508 (2008).

[5] W. Wang, H. Sukegawa, R. Shan, S. Mitani, and K. Inomata, Appl. Phys. Lett. **95**, 182502 (2009).

[6] N. Tezuka, N. Ikeda, F. Mitsuhashi, and S. Sugimoto, Appl. Phys. Lett. **94**, 162504 (2009).

[7] K.Z. Suzuki, R. Ranjbar, J. Okabayashi, Y. Miura, A. Sugihara, H. Tsuchiura, and S. Mizukami, Sci. Rep. **6**, 30249 (2016).

[8] J. Jeong, Y. Ferrante, S.V. Faleev, M.G. Samant, C. Felser, and S.S.P. Parkin, Nat. Commun. **7**, 10276 (2016).

[9] M. Yoshikawa, E. Kitagawa, T. Nagase, T. Daibou, M. Nagamine, K. Nishiyama, T. Kishi, and H. Yoda, IEEE Trans. Magn. **44**, 2573 (2008).

[10] G. Yang, D.L. Li, S.G. Wang, Q.L. Ma, S.H. Liang, H.X. Wei, X.F. Han, T. Hesjedal, R.C.C. Ward, A. Kohn, A. Elkayam, N. Tal, and X.-G. Zhang, J. Appl. Phys. **117**, 083904 (2015).

[11] P.G. Mather, J.C. Read, and R.A. Buhrman, Phys. Rev. B **73**, 205412 (2006).

[12] C. Wang, A. Kohn, S.G. Wang, L.Y. Chang, S.-Y. Choi, A.I. Kirkland, A.K. Petford-Long, and R.C.C. Ward, Phys. Rev. B **82**, 024428 (2010).

[13] H. Sukegawa, Y. Miura, S. Muramoto, S. Mitani, T. Niizeki, T. Ohkubo, K. Abe, M. Shirai, K. Inomata, and K. Hono, Phys. Rev. B **86**, 184401 (2012).

[14] H. Sukegawa, K. Inomata, and S. Mitani, Appl. Phys. Lett. **105**, 092403 (2014).

[15] T. Scheike, H. Sukegawa, K. Inomata, T. Ohkubo, K. Hono, and S. Mitani, Appl. Phys. Express **9**, 053004 (2016).

[16] M. Belmoubarik, H. Sukegawa, T. Ohkubo, S. Mitani, and K. Hono, Appl. Phys. Lett. **108**, 132404 (2016).

[17] A. Duluard, C. Bellouard, Y. Lu, M. Hehn, D. Lacour, F. Montaigne, G. Lengaigne, S. Andrieu, F. Bonell, and C. Tiusan, Phys. Rev. B **91**, 174403 (2015).

[18] Koji Tsunekawa, David D. Djayaprawira, Motonobu Nagai, Hiroki Maehara, Shinji Yamagata, Naoki Watanabe, Shinji Yuasa, Yoshishige Suzuki, and Koji Ando, Appl. Phys. Lett. **87**, 072503 (2005).

[19] H. Sukegawa, H. Xiu, T. Ohkubo, T. Furubayashi, T. Niizeki, W. Wang, S. Kasai, S. Mitani, K. Inomata, and K. Hono, Appl. Phys. Lett. **96**, 212505 (2010).

[20] T. Niizeki, H. Sukegawa, S. Mitani, N. Tezuka, and K. Inomata, Appl. Phys. Lett. **99**, 182508 (2011).

[21] Y. Miura, S. Muramoto, K. Abe, and M. Shirai, Phys. Rev. B **86**, 024426 (2012).

[22] G.X. Miao, Y.J. Park, J.S. Moodera, M. Seibt, G. Eilers, and M. Münzenberg, Phys. Rev. Lett. **100**, 246803 (2008).

[23] V. Drewello, M. Schäfers, O. Schebaum, A.A. Khan, J. Münchenberger, J. Schmalhorst, G. Reiss, and A. Thomas, Phys. Rev. B **79**, 174417 (2009).

# Captions

**Fig. 1:** (a) Schematic presentation of the Fe/MgAl$_2$O$_4$/Fe MTJ structure with a wedge-shaped Mg-Al insertion.

**Fig. 2:** (a) $t_{MgAl}$ dependence of the TMR ratio and $RA_P$ product for Fe/Mg-Al ($t_{MgAl}$)/MgAl$_2$O$_4$ (1.9 nm)/Fe MTJs (See the text for the meaning of the shaded I, II, and III regions). (b) TMR ratios as a function of the magnetic field for Fe/Mg-Al ($t_{MgAl}$ = 0.02, 0.22 and 0.45 nm)/MgAl$_2$O$_4$ (1.9 nm)/Fe MTJs. These data were measured at RT using 10 µA dc current (bias voltage < 10 mV).

**Fig. 3:** (a) Bias voltage dependence of normalized TMR ratio and (b) $t_{MgAl}$ dependence of $V_{half}$ values of Fe/Mg-Al ($t_{MgAl}$)/MgAl$_2$O$_4$ (1.9 nm)/Fe MTJs. The + (−) signs of $V_{half}$ corresponds to the applied positive (negative) bias voltage as defined in the upper side schemes. For (b), the regions I, II and III are illustrated for comparison.

**Fig. 4:** (a) Bias voltage dependence of (a) $G_P/G_P(0\ V)$ and (b) $G_{AP}/G_{AP}(0\ V)$ spectra of Fe/Mg-Al ($t_{MgAl}$)/MgAl$_2$O$_4$ (1.9 nm)/Fe MTJs measured at RT. Inset of (b) is the bias voltage dependence of the calculated $dG_{AP}/dV$ for three $t_{MgAl}$ values.

# Fig. 1

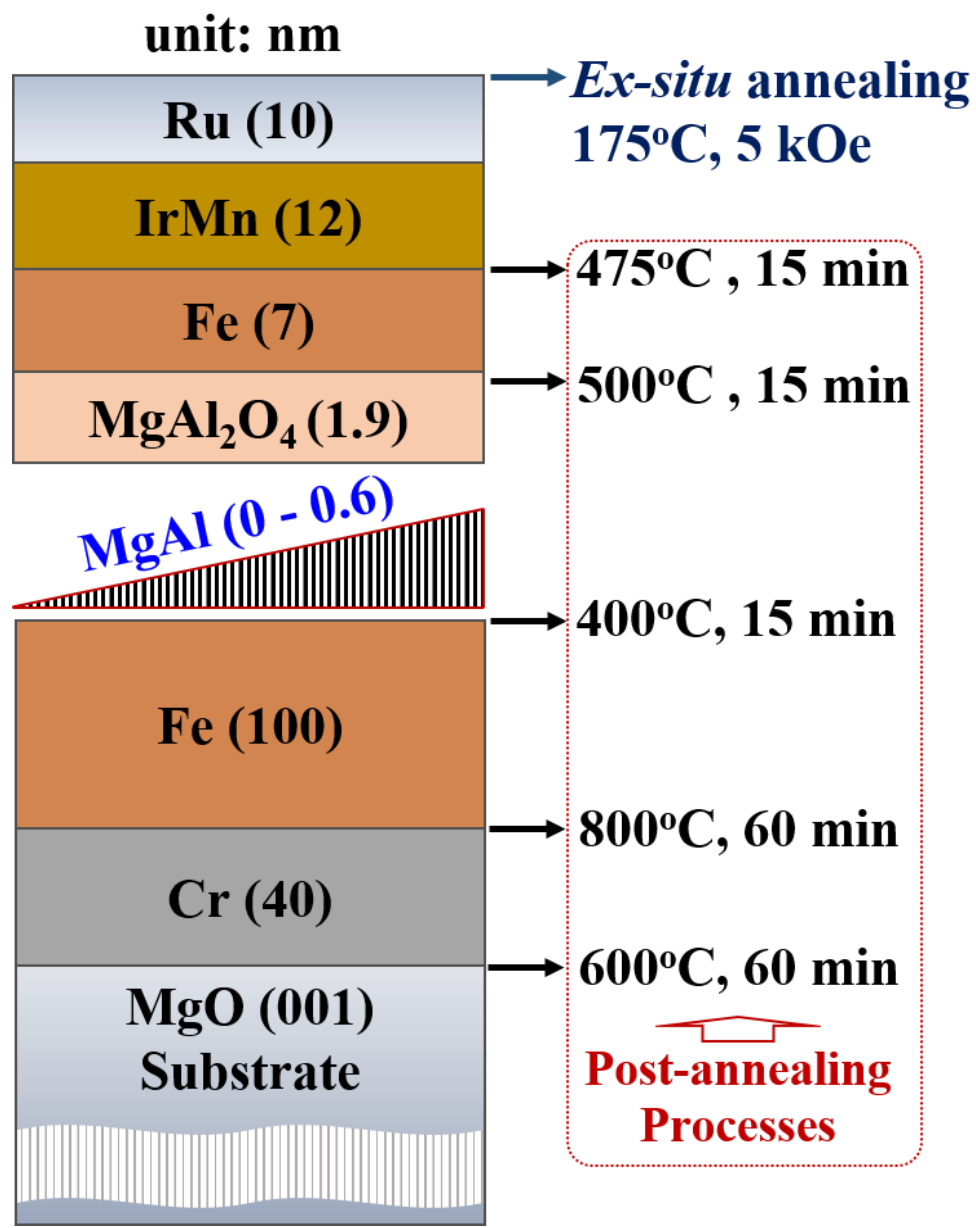

**Fig. 2**

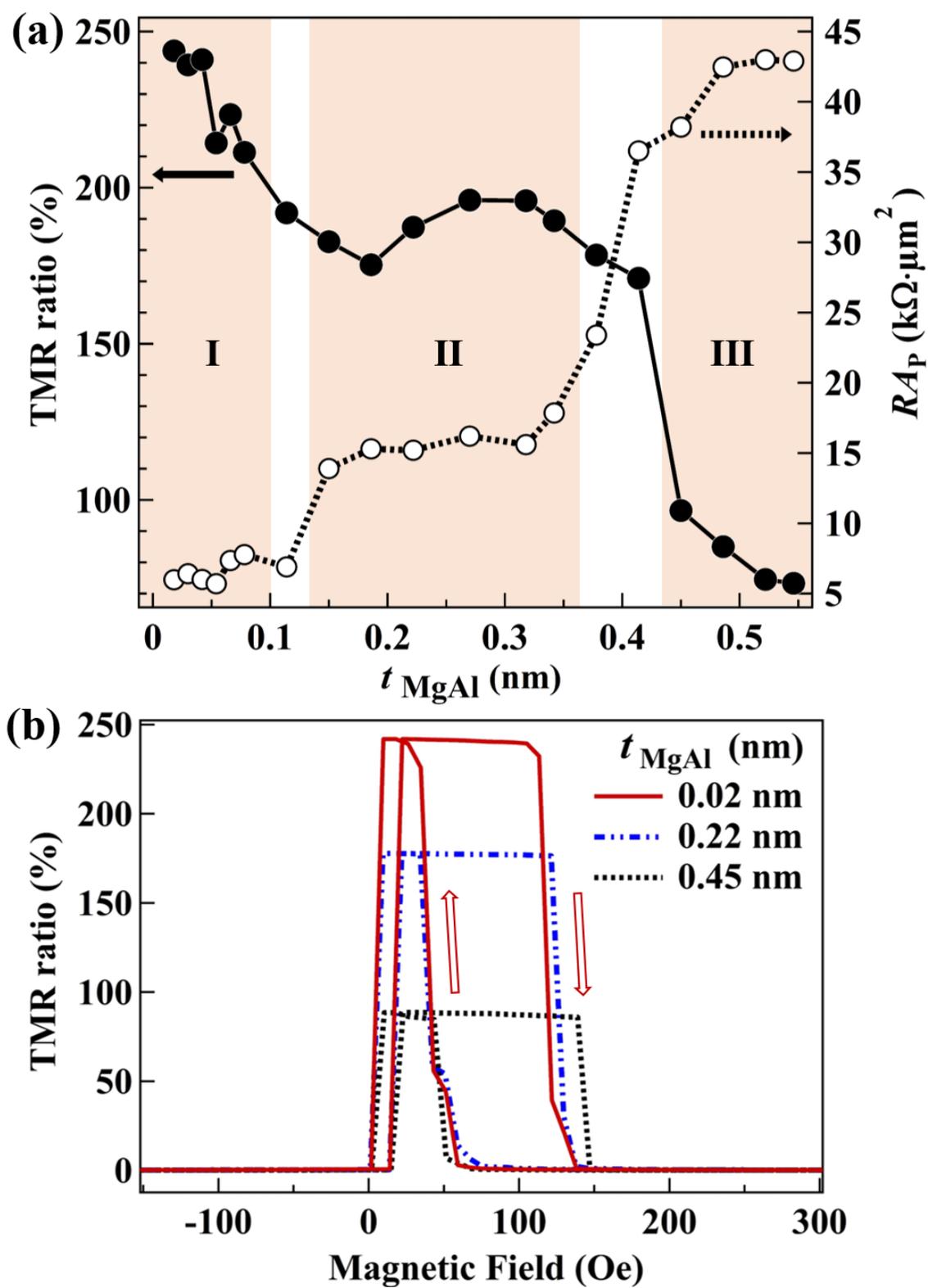

Fig. 3

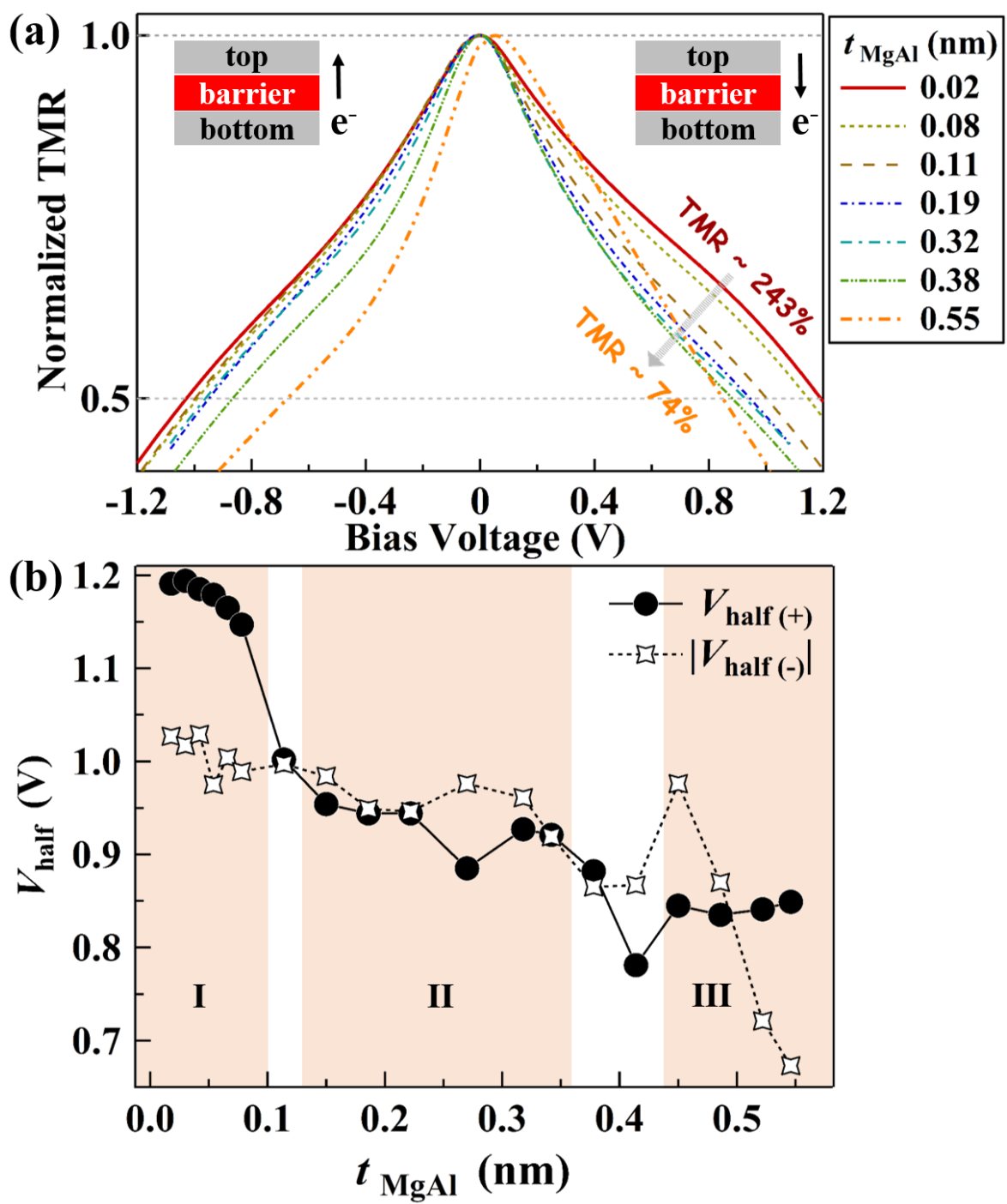

**Fig. 4**

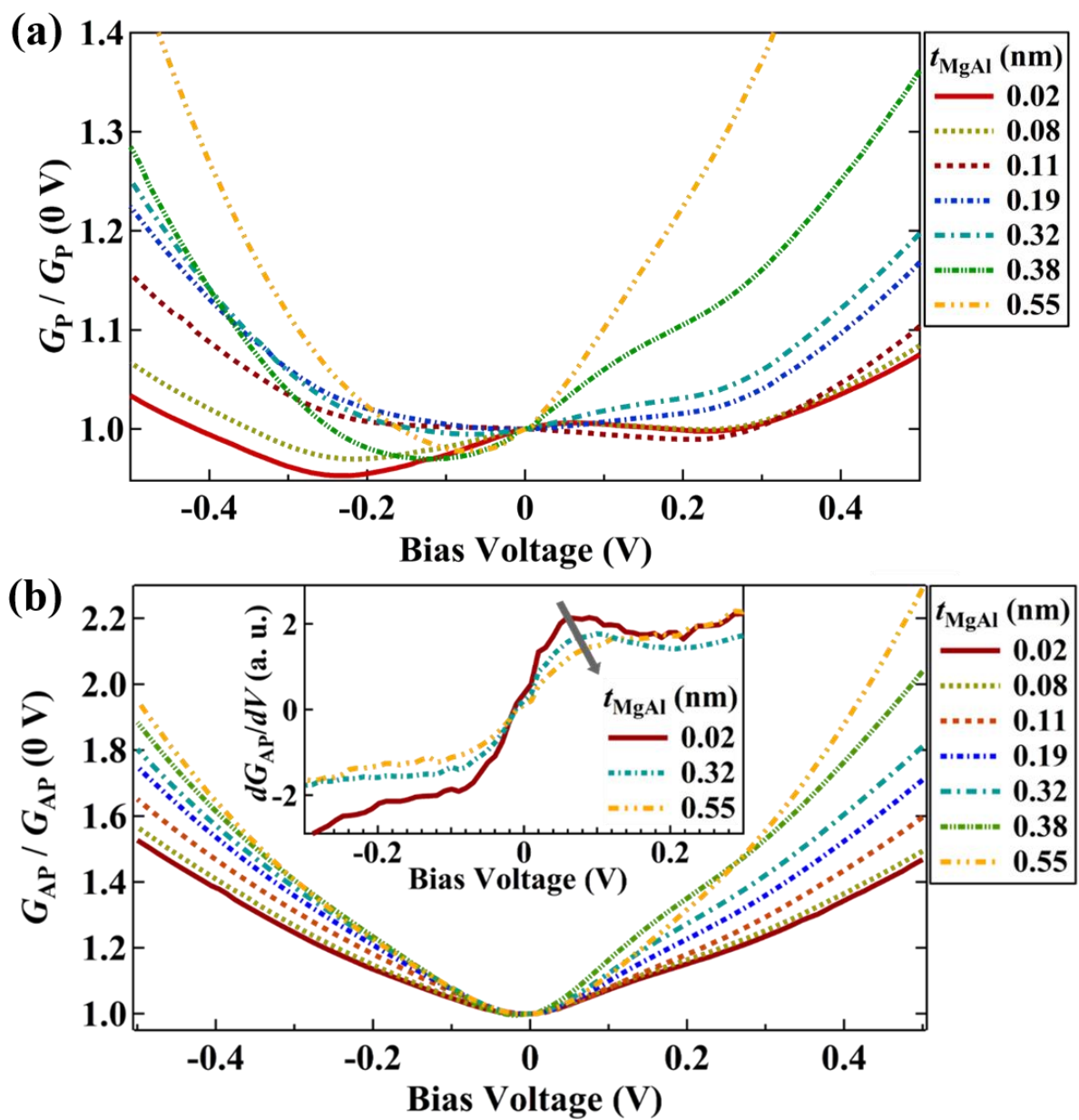